\documentclass{article}

\usepackage{makecell}

\usepackage{PRIMEarxiv}
\usepackage[autostyle]{csquotes} 
\usepackage{float}
\usepackage{natbib}
\usepackage{amsmath}
\usepackage{ntheorem}
\theoremseparator{:}
\newtheorem{hyp}{Hypothesis}
\usepackage{setspace}
\usepackage[utf8]{inputenc} % allow utf-8 input
\usepackage[T1]{fontenc}    % use 8-bit T1 fonts
\usepackage{hyperref}       % hyperlinks
\usepackage{url}            % simple URL typesetting
\usepackage{booktabs}       % professional-quality tables
\usepackage{amsfonts}       % blackboard math symbols
\usepackage{nicefrac}       % compact symbols for 1/2, etc.
\usepackage{microtype}      % microtypography
\usepackage{lipsum}
\usepackage{fancyhdr}       % header
\usepackage{graphicx}       % graphics
\graphicspath{{media/}}     % organize your images and other figures under media/ folder

\title{The Effect of Crypto Rewards in Fundraising: From a Quasi-Experiment to a Dictator Game
}

\author{
  Xue (Jane) Tan \\
  Information Technology \& Operations Management \\
  Southern Methodist University \\
  Dallas, TX 75275\\
  \texttt{janetan@smu.edu} \\
   \And
  Yong Tan \\
  Foster School of Business \\
  University of Washington \\
  Seattle, WA 98195\\
  \texttt{ytan@uw.edu} \\ 
  }
\begin{document}
\maketitle

\begin{abstract}
Conditional thank-you gifts are one of the most widely used incentives for charitable giving. Past studies explored non-monetary thank-you gifts (e.g., mugs and shirts) and monetary thank-you gifts (e.g., rebates that return some of the donations to the giver). Following the rapid growth of blockchain technology, a novel form of thank-you gifts emerged: the crypto rewards. Through two studies, we analyze crypto thank-you gifts to shed light on fundraising designs in the digital world. In Study I, we examine the Ukrainian government's crypto fundraising plea that accepts donations in both Ethereum and Bitcoin. We find that Ethereum is substantially more effective in enticing giving than Bitcoin, as the hourly donation count increased 706.07\% more for Ethereum than for Bitcoin when crypto rewards are present. This is likely because the crypto rewards are more likely to be issued on Ethereum than Bitcoin. However, the decrease in contribution sizes is also more substantial in Ethereum than in Bitcoin in response to the crypto rewards. In Study II, we conducted a laboratory experiment following a dictator game design to investigate the impact of crypto rewards in a more general scenario, with the crypto rewards specified as non-fungible tokens (NFTs). The crypto rewards in Study II carry no monetary value but only serve to recognize donors symbolically. As such, the NFT thank-you gifts did not effectively induce people to donate; a traditional 1:1 donation matching strictly outperforms both the condition without thank-you gifts and the condition with NFT thank-you gifts. Nevertheless, the NFT thank-you gifts effectively increased the contribution sizes, conditional on the choice to give, when the NFT's graphic design primes donor identity and encompasses the charity recipient.
%that accepts Ether (i.e., the currency of the Ethereum blockchain) and Bitcoin (i.e., the currency of the Bitcoin blockchain) over a seven-day period, we analyze the impact of crypto rewards on fundraising performance. Crypto rewards are newly minted tokens that are usually valueless initially and grow in value if the corresponding cause is well received. By estimating the ordered treatment effects in a modified difference-in-differences model, we find that the crypto rewards lead to an 812.48\% stronger donation count increase for Ethereum than for Bitcoin, given that the crypto rewards are more likely to be issued on the Ethereum blockchain, which has higher programmability to support smart contracts. In terms of the donation amount, we find a 30.1\% stronger decrease in the average donation amount from Ethereum for small donations ($\leq \$250$); the rewards pose similar impacts on the average donation size for the two blockchains for large donations ($>\$250$). We then probed into Ethereum donation transactions and discovered the positive moderating effects of (1) Ethereum Name Service (ENS) adoption, (2) intermediary platform (e.g., Coinbase) usage, and (3) NFT asset holdings. Our analyses generate rich implications for the crypto community, fundraising organizations, and policymakers. 
\end{abstract}

% keywords can be removed

\section{Introduction}

Thank-you gifts are conditional gifts, also named donor appreciation gifts in practice or donor premiums in economics, offered by fundraisers to donors. These gifts are a prevalent feature in fundraising activities, with 1\%-2\% of the fundraising revenue allocated to thank-you gifts \citep{recognitionartEstablishingBudgetfor}. Previous research has predominantly examined non-monetary thank-you gifts, such as mugs, tote bags, and shirts \citep{newman2012counterintuitive}, as well as monetary gifts, such as rebates \citep{list2008introduction}. This study investigates a novel form of thank-you gift: crypto rewards, which are represented by immutable fungible and non-fungible tokens (NFTs) to crypto donors. 

Crypto rewards deserve scholarly attention due to their unique position as hybrids of non-monetary and monetary thank-you gifts. Like non-monetary thank-you gifts, they enable individuals to signal their prosociality both to themselves and to others, leveraging the immutable nature of blockchain technology. Particularly when crypto rewards are in the form of NFTs, they serve to publicly showcase the achievements, milestones, and social standing of their owners within the digital realm. Meanwhile, crypto rewards possess monetary or investment value. Although they are usually valueless when they are initially minted, their value could increase when they are exchanged if the associated causes gain widespread support. This investment characteristic distinguishes crypto rewards from traditional monetary thank-you gifts like rebates.

%An NFT thank-you gift recognizes one's good deeds, priming the donor's online persona as an altruistic being. It also serves as a digital signature to indicate that the donor supports the fundraiser's endeavors.  

Crypto donors are becoming increasingly significant in the realm of fundraising. With 420 million cryptocurrency users in 2023, and projections indicating a rise to 1 billion users by 2030 \citep{thegivingblock2023Annual}, crypto donors represent a burgeoning and vital demographic for charitable giving. The Giving Block, which provides solutions for cryptocurrency donations within the nonprofit sector, reports that more than 150 nonprofits are able to raise over \$100 million of annual revenue from crypto donors, as crypto fundraising allows nonprofits to access a distinct demographic -- younger, wealthier, and more dominated by male -- that traditional fundraising channels often fail to reach \citep{thegivingblock2023Annual}.

The following examples demonstrate the surging prevalence of crypto rewards. The first example is Giveth (https://giveth.io), a blockchain-based crowdfunding platform that has raised \$2,239,120 for 2,875 projects from 6,646 givers at the time of writing. Giveth differs from traditional crowdfunding in that it rewards givers with GIV tokens when they support verified projects. The GIV tokens can be used by token holders to influence the roadmap and mission of the Giveth ecosystem through a voting mechanism. Moreover, Giveth offers NFTs as thank-you gifts for every contribution of \$100 or more to the Givth platform. \footnote{https://giveth.io/nft/mint} The second example is UkraineDAO, a decentralized autonomous organization (DAO) that raises funds to help Ukraine defend itself. UkraineDAO offers the LOVE tokens to its crypto donors as thank-you gifts. These love tokens could translate into the collective ownership of a Ukraine flag NFT. The Ukrainian flag NFT was sold for 2,258 ETH (about \$6.75 million as of then),\footnote{https://www.coindesk.com/tech/2022/03/02/ukrainian-flag-nft-raises-675m-for-countrys-war-efforts/} and donors could profit from the sales.

This study aims to explore the potential of blockchain technology for social good by examining the impact of crypto rewards on fundraising performance with the following questions: 
\begin{itemize}
  \item RQ1: Are crypto rewards effective in improving fundraising performance?
  \item RQ2: Would crypto rewards be more effective than donation matching (i.e., adding extra donations to the giver's contributions)?
  \item RQ3: When crypto rewards are in the format of NFTs, what graphic designing factors of the NFTs would impact the fundraising performance?
\end{itemize}

The concept of providing donors with crypto rewards draws inspiration from the ``airdrop'' in the initial coin offering (ICO) model, which serves as a promotional strategy to increase awareness for blockchain-based projects \citep{li2021operation}. In ICOs, rewards are distributed to community members who engage in activities such as following a Twitter account or joining a Telegram group. These reward tokens are typically valueless at inception but can be utilized within blockchain-based projects to access services and goods upon their launch. In this study, we examine crypto rewards within a broader context, where crypto donations are directed towards social causes rather than exclusively supporting blockchain-based projects. The value of these crypto rewards is not tied to the success of a specific blockchain project but rather to the societal reception of the associated social cause. This application of blockchain technology warrants promising potential for social good, expanding the impact of blockchain technology beyond the confines of the blockchain industry.

We use a mixed-method approach with two studies to examine the impact of crypto rewards as this approach is most appropriate for studying new technologies \citep{velichety2019quality}. Study I is based on the Ukrainian government's crypto fundraising plea, which accepts donations from both Ethereum and Bitcoin. Donors who contribute via Ethereum are more likely to receive the airdrop than those who contribute via Bitcoin due to Ethereum's stronger programmability to support smart contracts. Indeed, while donation trends of these two cryptocurrencies were similar prior to the announcement of the airdrop, the donation trends diverged drastically after the announcement of the airdrop. We performed both blockchain-level and transaction-level analyses to understand the impact of this airdrop, following a difference-in-differences approach \citep{fricke2017identification}. At a blockchain level, We find that hourly donation counts increased 706.07\% more for Ethereum than for Bitcoin in response to the airdrop. Further, the average contribution size for Ethereum dropped 57\% more than that for Bitcoin following the airdrop. At a transaction level, we performed Coarsened Exact Matching to identify comparable wallets from Ethereum and Bitcoin based on time since the first transaction and historical transaction volumes. We continue to find that the contribution size dropped more aggressively for Ethereum than for Bitcoin in response to the airdrop. We further performed moderation analyses on Ethereum transactions to show that the negative link between crypto rewards and contribution size is lessened when the wallet is registered with Ethereum Name Service (ENS) and when the donation is transacted on intermediary platforms. 

While Study I highlights the significant potential of using airdrops to promote social causes in crypto fundraising, replicating this success may be challenging due to the uniquely high visibility of the event. In Study II, we assess the impact of crypto rewards in a more general scenario by conducting a dictator game,  ``a celebrated workhorse of experimental economics and social psychology'' to understand charitable giving \citep{cartwright2023using}. We recruited 268 subjects from Prolific, and the screening requirement was that the subjects must be owners or ex-owners of NFTs. Following the protocol of dictator games \citep{engel2011dictator}, subjects first work on two copyediting tasks to earn a flat-rate income of \$2 on top of their participation fee (\$1). They are then asked to allocate the \$2 between themselves and an international charity, Doctors without Borders. We use a between-subject design and assign these subjects into five groups: a baseline group without a thank-you gift, a donation matching group that receives a 1:1 matching offer to further support Doctors without Borders, and three NFT reward groups with different NFT designs. Unlike the airdrops in Study I, the NFT rewards in Study II were created by the experimenters and likely offered only symbolic recognition but not monetary value. Consequently, we found no significant effect of these NFT rewards on the decision to donate. Consistent with previous research demonstrating the stronger impact of donation matching compared to rebates \citep{eckel2003rebate, eckel2006subsidizing}, the matching group was the most effective in encouraging donations. However, the design of NFTs significantly influenced the amount donated among those who chose to give. Specifically, donors who perceived donor identity as relevant to themselves (as discussed in \cite{kessler2018identity}) were likely to donate more when the reward NFT both highlighted the donor's identity and included the recipient's identity. This suggests that the alignment of the NFT design with donor identity can enhance the effectiveness of crypto rewards.

This study makes unique contributions to the literature on charitable fundraising. On the one hand, regarding whether to donate, the differential findings of Studies I and II highlight the critical requirement for a crypto reward to be effective -- crypto rewards could successfully entice the choice to give when they are monetarily appealing to prospective donors. This is in contrast to the detrimental effects of extrinsic motives for charitable giving suggested by past studies \citep{frey2001motivation,frey1997cost}. Our findings underscore the crypto community's strong investment mindset, as the monetary value of crypto rewards varies with the public reception of the cause and the stake of the entity that initiated the fundraising campaign. On the other hand, we show in Study II that the design of NFTs could serve the role of donor identity prime to increase the gift size conditional on giving, uncovering the multi-faceted roles of NFT rewards and contributing to the studies that examined the choices and framing of thank-you gifts \citep{zlatev2016selfishly}. We also directly contribute to \cite{chao2017demotivating} and \cite{chao2022self}, who discovered that visually salient thank-you gifts would reduce giving by shifting people's attention to focus on self-interest rather than their intrinsic motives. In great contrast, the visual display of NFTs offers a novel avenue to incentivize giving. Last but not least, we contribute to the large stream of literature on charitable giving based on dictator games. Ever since Daniel Kahneman performed the first dictator game over three decades ago, hundreds of papers have been published based on dictator games to understand various factors that affect individuals' decisions regarding giving (e.g., incentives, social factors, distributive concerns, framing, social distance, and demographics) \citep{engel2011dictator}. This study contributes to this literature with the inclusion of unique features of crypto rewards used in crypto fundraising campaigns.  

\section{Theoretical Development}
\label{sec:theory}

\subsection{Self-interest and Charitable Giving}
The nature of crypto rewards is a ``thank-you'' gift for making a crypto donation. Conditional thank-you gifts are ubiquitous extrinsic incentives used in charitable fundraising, where donors get a non-monetary gift (e.g., mugs, tote bags, and shirts) or a monetary gift (e.g., rebate) conditional on their charitable contribution \citep{newman2012counterintuitive, eckel2006subsidizing}. The effect of extrinsic incentives on prosocial behavior has been widely studied in the literature of economics, psychology, and information systems \citep{newman2012counterintuitive,gneezy2000fine,liu2021does, eckel2006subsidizing, list2002effects}. Prior works identify the facilitating role of extrinsic motivations in giving because humans are primarily motivated by self-interest \citep{kohn2008brighter}. Self-interest has become a social norm to the extent that people would be hesitant to donate to a charitable cause even when they have strong feelings of compassion for it \citep{miller1994collective,miller1999norm}. In such cases, extrinsic incentives could work as an ``excuse'' to rationalize people's prosocial behavior by concealing their prosocial motivations \citep{holmes2002committing}. Extrinsic motives could also manifest through reciprocity, where the thank-you gifts activate donors' feelings of reciprocity such that they give more \citep{briers2007adding}. As evidence, \cite{falk2007gift} finds that the frequency of donations increased by 17\% when a small gift was given to donors and by 75\% when a large gift was given. 

On the contrary, people could reduce giving when extrinsic incentives are provided because extrinsic motivations would crowd out intrinsic motivations \citep{frey1997cost,frey2001motivation,gneezy2000fine,benabou2006incentives,chao2022self}. Differing from extrinsic motivations, which are activated by monetary rewards, praise, or fame, intrinsic motivations are related to activities that people undertake because they derive satisfaction from them. As \citet[p.746]{frey1997cost} stated, \enquote{If a person derives intrinsic benefits simply by behaving in an altruistic manner or by living up to her civic duty, paying her for this service reduces her option of indulging in altruistic feelings.} Many studies discover evidence in support of the motivation crowding-out theory. \cite{newman2012counterintuitive} find that among the donors who are willing to contribute, those who were offered a thank-you gift donated a significantly lower amount than those who were not offered a thank-you gift. \cite{chao2017demotivating} further find that the negative effect of extrinsic motivation from a thank-you gift is only present when the gift is visually salient to occupy the prospective donor's attention.

Despite the opposing predictions, we propose that self-interest dominates the decision of crypto donations because investment is key to blockchain users \citep{kim2020comparison}.

\begin{hyp}  \label{hyp:first}
Crypto rewards positively impact charitable giving, especially when the crypto rewards could offer potential returns.
\end{hyp}

\subsection{Donation Matching}
The next hypothesis is designed to enhance the understanding of crypto rewards by drawing a comparison with the fundraising strategy of donation matching. Donation matching has been identified as one of the most effective means of fundraising. It refers to the practice that a large donor (e.g., an employer or a charitable foundation) matches individuals' contributions to a specific cause to increase the gift. Eckel and Grossman showed in both within-subject and between-subject lab experiments that a donation match is significantly more effective than a rebate (returning a portion of the donation to the giver) in fundraising performance \citep{eckel2006subsidizing,eckel2003rebate}. \cite{gandullia2018price} performed online experiments to show similar findings that donation matching is more effective than rebates. \cite{kamas2010can} showed that donation matching is effective even for self-interested donors (as compared to social surplus maximizers and inequity averters). The studies that compared donation match with rebates controlled for the value of the money added to the gift or returned to the donor. In our context, the value of crypto rewards is ambiguous and key to the relative advantage of these two fundraising strategies.

As discussed, the nature of crypto thank-you gifts is multifaceted. On the one hand, crypto rewards resemble monetary thank-you gifts, such as rebates, by stimulating giving through self-interest. On the other hand, they function similarly to non-monetary thank-you gifts, like mugs, tote bags, and shirts, by reinforcing acts of giving through symbolic recognition. When crypto rewards offer high monetary returns or strong symbolic recognition, they can be more appealing than donation matching. Conversely, when the monetary return of the crypto rewards and the symbolic recognition are low, they may not be as effective as donation matching.

\begin{hyp}  \label{hyp:second}
Crypto rewards are less effective than a donation match in fundraising when the crypto rewards offer low monetary returns or symbolic recognition and are more effective otherwise.
\end{hyp}

\subsection{Donor Identity Prime} 
As we mentioned, crypto rewards not only carry monetary value but also could reinforce one's self-identity and public identity as an altruist. Identity refers to a person's sense of self that is ``associated with different social categories and how people in these categories should behave \citep[p.~715]{akerlof2000economics}.'' \cite{kessler2018identity} find from field experiments run by the American Red Cross that appeal that prime individuals' identity as previous givers results in more donations. This is because people tend to adjust their behaviors such that their behaviors will match the norms or prescriptions associated with their identity \citep{akerlof2000economics}. As discussed in \cite{kessler2018identity}, a large stream of works has demonstrated the power of priming -- even remarkably small environmental cues could change which facet of individual identity is salient at a certain point \citep{steele1997threat}. 

Other than priming donor's identity, it is generally believed that disclosing the recipient's identity (through logos) could enhance the value of the thank-you gifts. As evidence, \cite{jung2014paying} considered the value of the Cal logo/UCB affiliation in the thank-you gift of mugs, where UCB stands for University of California, Berkeley.

We propose that the graphic design of NFT gifts plays a role in the decision to give. The design that primes donor identity and the recipient's identity, as reflected in logos, would increase giving because the former signals prosociality, and the latter demonstrates one's commitment to the corresponding endeavor. Since the crypto community values decentralization, freedom, and democracy, cryptocurrency holders likely desire to hold the crypto reward tokens that showcase their support for causes that align with their group identity \citep{ramaswamy_2022}. 

\begin{hyp}  \label{hyp:third}
The graphic design of NFT thank-you gifts that primes donor identity and the identity of the recipient positively influences crypto donations.
\end{hyp}

\section{Study I - A Quasi-experimental Study}
Study I is designed to test the hypothesis of H\ref{hyp:first} based on a Ukrainian crypto fundraising campaign.
\subsection{Context and Data Collection}
\begin{figure}
\begin{center}
\includegraphics[height=3.3in]{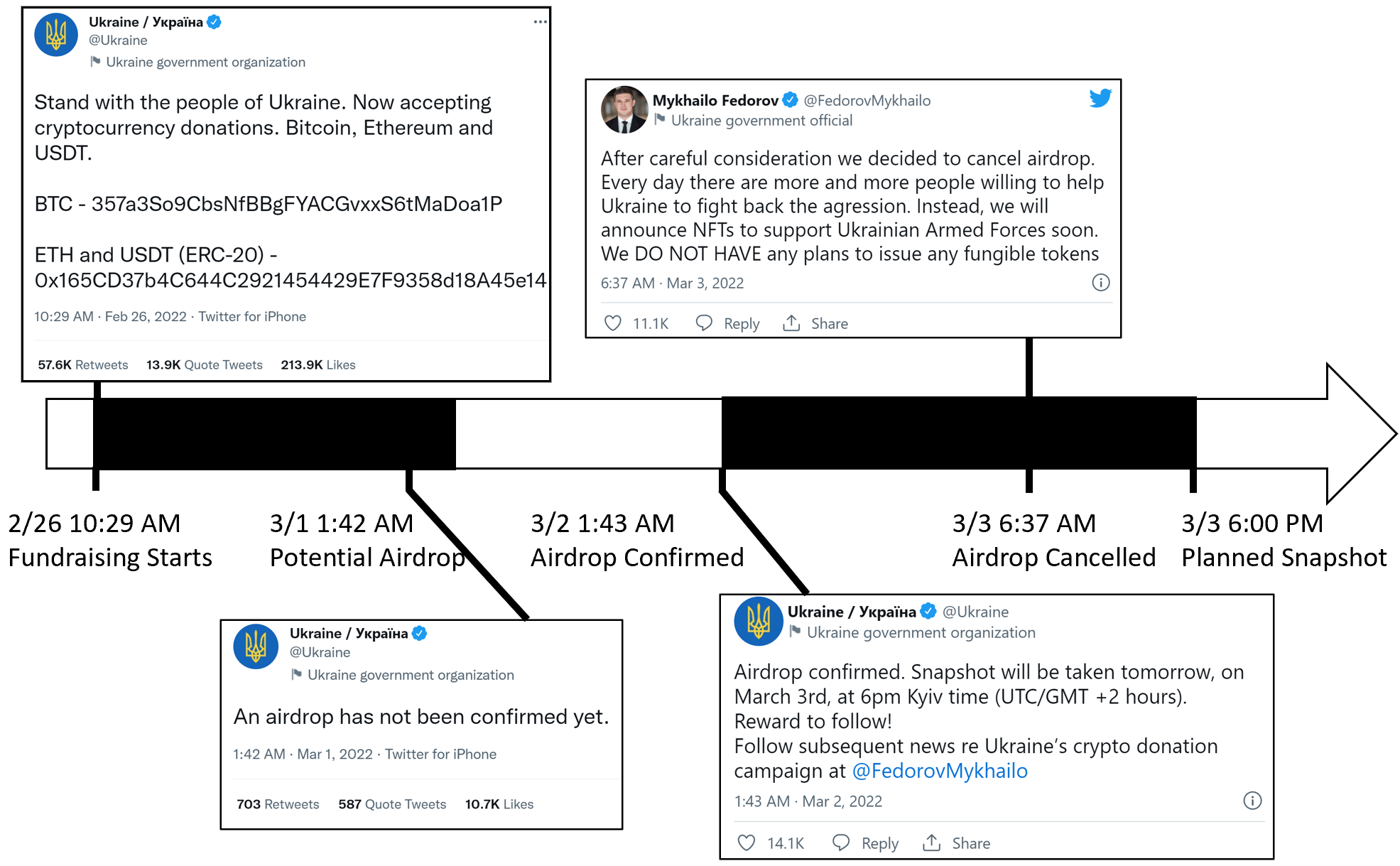}
\caption{Timeline} 
\label{fig:timeline}
\end{center}
\end{figure}

The Ukrainian government posted pleas for cryptocurrency donations on Feb. 26 at 10:29 AM, 2022 (UTC). Since Ukraine's banking system was at risk of a Russian attack, crypto offered an alternative financial structure to support Ukraine because it uses cryptography to secure transactions. This fundraising channel is different from other fundraising efforts made by nonprofit organizations because all the funds would be directly received by the Ukrainian government, avoiding overheads. In a tweet, the Ukrainian government announced their Ethereum and Bitcoin wallet addresses. Ether (ETH) is the native currency traded on the Ethereum blockchain, and Bitcoin (BTC) is the currency traded on the Bitcoin blockchain; both ETH and BTC are digital currencies based on the distributed ledger technology of blockchain. On March 1 at 1:43 AM, Ukraine announced that an ``airdrop" has not been confirmed, but formally announced on March 2 at 1:43 AM that they would reward donors who supported Ukraine with an airdrop. The planned snapshot of the list of donor wallet addresses would be taken the next day. While the initial announcement that ``An airdrop has not been confirmed yet" does not officially start the airdrop, people may react to this potential airdrop even before the official announcement is posted. One day later, on March 3 at 6:37 AM, the vice prime minister of Ukraine and the Minister of Digital Transformation of Ukraine announced the cancellation of this airdrop soon before the scheduled snapshot. The timeline is summarized in Figure \ref{fig:timeline}. \footnote{From the Ethereum wallet data we collected, the Ukrainian government issued NFT rewards to some Ethereum donors post campaigns despite of this cancellation.} 

We collected donation transactions between 1:00 AM on Feb. 26, 2022 and 6:00 PM on Mar.3, 2022, from the public wallets of Ukraine to focus on the airdrop. We calculated the USD value of donation contributions using the historical prices of Bitcoin and Ethereum based on the daily opening prices. There were 14,903 donation transactions on the Bitcoin blockchain and 69,709 donation transactions on the Ethereum blockchain during this observation window. In total, \$9,874,757 was raised from the Bitcoin blockchain, and \$16,043,036 was raised from the Ethereum blockchain. We excluded extreme donations above the 99.9th percentile (\$7727.13). We sum up donations from the same wallet if they are transacted within the same minute. We further removed transactions between 2:00 AM on Mar. 1, 2022 and 2:00 AM on Mar. 2 because this period is associated with a potential airdrop. We are left with 67,615 donation transactions.
\subsection{Identification}

The introduction of an airdrop as a reward to crypto donors offers a quasi-experiment for us to understand the impacts of the crypto rewards on donation counts and amounts. While the publicity of the fundraising event increased sharply after the announcement of the crypto rewards due to substantial media coverage, this temporal effect equivalently applies to both ETH and BTC because the wallet addresses for both ETH and BTC were included in the same tweet. However, the impact of the crypto rewards is likely much stronger on ETH than BTC because only ETH supports smart contracts, and most airdrops were issued through ETH. We perform a modified DiD analysis to exploit the difference between the impacts of the airdrop on Bitcoin and Ethereum in order to explore the causal impacts of crypto rewards \citep{wing2018designing}. This method has the same functional form as classic DiD but offers different interpretations based on the differential impacts of treatment with varying intensities \citep{fricke2017identification}. \cite{duflo2001schooling} used this method to understand the effect of school construction on schooling and labor market outcomes by comparing regions with low and high levels of newly constructed schools. \cite{felfe2015can} leveraged the regional variation in childcare expansion rates to understand the effect of formal childcare on maternal employment as well as child development. This extension of DiD requires not only the common trend assumption but also an \textit{equal effect size} assumption, which posits that users in ETH and BTC would respond similarly when high-intensity treatment (a higher likelihood of winning the crypto rewards) was imposed on the corresponding blockchain. In Section 1 of the Appendix, we discuss the identification strategy and assumptions in greater detail. 

We perform both an aggregated analysis (Section \ref{sec:agg}) and a transactional analysis (Section \ref{sec:tra}) to estimate the ordered treatment effect of the airdrop. The aggregated analysis provides causal inference for the hourly donation count; the transactional analysis offers insights into contribution sizes after accounting for the potential selection process.

\subsubsection{Aggregated Analysis at a Blockchain Level.}
\label{sec:agg}
At a blockchain level, the econometric model we estimate is specified as below ($c$ denotes the blockchain and $t$ denotes the hours):
\begin{equation}
Outcome_{c,t}=\beta_{0}+\beta_{1} Ether_{c} \times Airdrop_{t}
+\beta_{2}Ether_{c}+\beta_{3}Airdrop_{t}+\beta_{4}FeeRate_{c,t}+\eta_{t}+\epsilon_{c,t},
\end{equation}
where the dependent variable $Outcome_{c,t}$ can be operationalized as two aggregated measures: the logarithm of the number of hourly donations ($DonationCount_{c,t}$) and the logarithm of the average contribution sizes ($AvgDonationSize_{c,t}$). These outcomes are both log-transformed after adding one because they are highly skewed. Our key independent variable, $Airdrop_{t}$, is a binary variable that takes the value of one if the airdrop has been announced but the snapshot has not been taken and zero otherwise. $Ether_{c}$ is a binary variable that takes the value of one if the currency is Ether and zero if it is Bitcoin. To identify the ordered treatment effect, we also include two-way fixed effects \citep{fricke2017identification}. We use $\eta_{t}$ to represent the hourly time dummy variables that account for the time-level fixed effects. The time effects could come from the dynamic situations in Ukraine, the increasing awareness of the crypto fundraising event, or simply donors' varying availability of time. Such temporal trends affect the Bitcoin and Ethereum blockchains in the same way. The systematic difference between BTC and ETH is represented by the group-level fixed effects, denoted as $Ether_{c}$. In addition, we account for the transaction fee rate ($FeeRate_{c,t}$) for both ETH and BTC using the transaction data on the Ethereum and Bitcoin blockchains from Google Big Query. For ETH, we calculate the average gas price at every point in time; for BTC, we calculate the ratio between fee and output value at every point in time. The coefficient of our interest is $\beta_{1}$ as it indicates the differential impacts of the airdrop on the blockchains of Bitcoin and Ethereum. For the aggregated data, we have 134 discrete observation points for both Ethereum and Bitcoin.

\subsubsection{Transactional Analysis at a Wallet Level.}
\label{sec:tra}
The aggregated analysis sheds light on the difference in donation behavior between the two blockchains when crypto rewards become available. However, it does not account for the selection process that individual donors go through when deciding whether to give via ETH or BTC. Specifically, people with a strong investment mindset may choose to donate via ETH to increase the likelihood of receiving crypto rewards in the first place. To mitigate the selection issue, we perform matching at a wallet level before examining the outcome of donation sizes ($DonSize_{w,t}$), where we use $w$ to denote the index for a wallet and $t$ for time. We no longer examine $AvgDonSize$ or $DonCount$, which are aggregated measures. The matching process proceeds as follows. First, for every donation made on BTC and ETH, we collected wallet information from BlockChair by using the paid APIs (https://blockchair.com/). Second, we removed all wallets that have exactly one receiving transaction and one sending transaction. These wallets were only used for this fundraising plea, and their owners are more likely to be driven by an opportunistic motive. Blockchair keeps the latest 100 transactions for every wallet, and we removed wallets whose latest 100 transactions did not include their donation to Ukraine. This allows us to exclude extremely active users, who are likely organizations rather than individuals; it also allows us to accurately construct the number of historical transactions for every wallet. Third, we removed ETH wallets that registered to use the Ethereum Name Service (ENS) because these accounts are less likely to be comparable with BTC wallets. ENS is a paid decentralized naming system that allows users to use human-readable names instead of hexadecimal characters to find Ethereum addresses. Then, we performed a wallet-level matching using Coarsened Exact Matching (CEM) based on (1) years since the first spending transaction ($TimeSinceFirstSend_{w,t}$), (2) years since the first receiving transaction ($TimeSinceFirstRec_{w,t}$), and (3) the logarithm of past transaction counts ($Log(PastTrans_{w}+1)$). After matching, we are left with 4,498 BTC transactions and 26,577 ETH transactions in 157 stratums and the corresponding weights were calculated to guarantee comparability. We achieved a good balance between treatment and control, as the standardized mean difference (SMD) for $TimeSinceFirstSend_{w,t}$ is 0.034, the SMD for $TimeSinceFirstRec_{w,t}$ is -0.0004, and the SMD for $Log(PastTrans_{w}+1)$ is -0.003. These SMDs are well below the threshold of 0.1. We then estimate the weighted linear regression model and incorporate the subclass to generate average treatment effects. Similar to the aggregated analysis, we included $Airdorp_{t}$ to represent whether the crypto rewards were available and $Ether_{w}$ to indicate whether the corresponding wallet is from Ethereum. We also included $FeeRate_{w,t}$, which is calculated at a minute granularity, hourly time dummies, and other variables we have used for the matching process to account for the time-varying features that could affect the contribution sizes. Based on the resulting matched samples, we estimate the following model:

\begin{equation} \label{eq1}
\begin{split}
DonSize_{w,t} & =\beta_{0}+\beta_{1} Ether_{w} \times Airdrop_{t}   \\
&+\beta_{2}Ether_{w}+\beta_{3}Airdrop_{t}+\beta_{4}FeeRate_{w,t}+\beta_{5}TimeSinceFirstSend_{w,t} \\
&+\beta_{6}TimeSinceFirstRec_{w,t}+\beta_{7}Log(PastTrans_{w}+1)+\eta_{t}+\epsilon_{w,t}.
\end{split}
\end{equation}

\subsection{Data}
In Table \ref{table:summarybygroup}, we summarize the aggregated data regarding $DonCount$ and $AvgDonSize$ in Panel A for Bitcoin and in Panel C for Ethereum. We summarize the transactional data in terms of $PastTransaction$ in Panel B for Bitcoin and in Panel D for Ethereum. 

\begin{figure}
    \centering
    \includegraphics[width=0.8\linewidth]{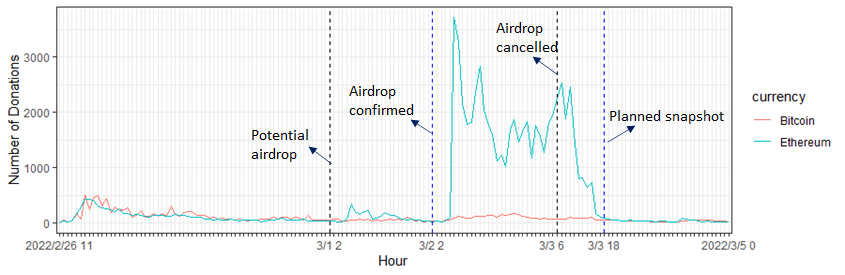}
    \caption{Donation Counts}
    \label{fig:donation_count}
\end{figure}

\begin{figure}
    \centering
    \includegraphics[width=0.8\linewidth]{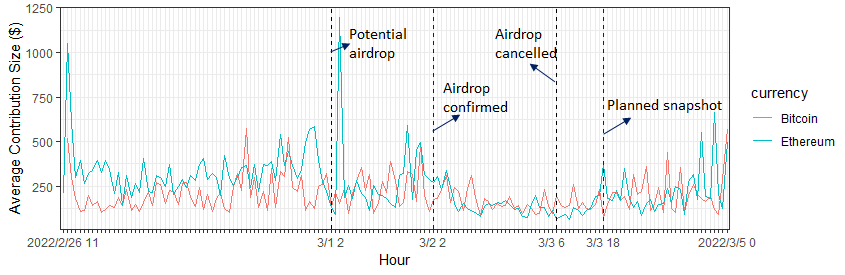}
    \caption{Average Donation Size}
    \label{fig:donation_avg}
\end{figure}

\begin{table}
\caption{Summary Statistics by Groups}
\renewcommand{\arraystretch}{1}
\begin{tabular}{llllllll}
\hline \hline
                   & \multicolumn{3}{c}{$Airdrop_{t}$=0} & \multicolumn{3}{c}{$Airdrop_{t}$=1}   & \multicolumn{1}{c}{Welch's $t$-test} \\ \cline{2-8} 
 &
  \multicolumn{1}{c}{Mean} &
  Median &
  \multicolumn{1}{c}{S.E.} &
  \multicolumn{1}{c}{Mean} &
  Median &
  \multicolumn{1}{c}{S.E.} &
  \multicolumn{1}{c}{$t$-stats} \\ \hline
\multicolumn{8}{l}{Panel A. Aggregated Contributions on Bitcoin ($Ether_{c}$=0)}                                                           \\ \hline
$DonCount_{c,t}$      & 106.75      & 71     & 104.92    & 86.83    & 84      & 37.82   & -1.60                               \\
$AvgDonSize_{c,t}$    & 180.65     & 161.85    & 72.90     & 152.32   & 141.35    & 49.62  & -2.60**                             \\
\hline
\multicolumn{8}{l}{Panel B. Transactional Contributions on Bitcoin ($Ether_{c}$=0)}                                                           \\ \hline

$PastTransaction_{c,i}$    & 147.52     & 11     & 2389.07     & 60.16    & 8	   & 624.24   & -1.24                         \\
\hline                         \\ 
 \hline 
\multicolumn{8}{l}{Panel C. Aggregated Contributions on Ethereum ($Ether_{c}$=1)}                                                         \\ \hline
$DonCount_{c,t}$      & 95.55      & 56     & 93.56   & 1215.73   & 1451    & 845.36   &  8.46***                       \\
$AvgDonSize_{c,t}$    & 267.16     & 261.96    & 109.32    & 132.70   & 129.53   & 46.51    & -9.98***    \\ \hline

\multicolumn{8}{l}{Panel D. Transactional Contributions on Ethereum ($Ether_{c}$=0)}                                                           \\ \hline

$PastTransaction_{c,i}$    & 101.52     & 30   & 629.82
    & 77.66  & 20  & 538.06    & -3.26***                                   \\
\hline
\hline 
\textit{Note:}  & \multicolumn{7}{r}{$^{*}$p$<$0.1; $^{**}$p$<$0.05; $^{***}$p$<$0.01} \\ 
\end{tabular}
\label{table:summarybygroup}
\end{table}
We performed Welch's t-test to compare the conditions with and without the airdrop because the samples in different groups have different variances. From Table \ref{table:summarybygroup}, we learn that the donation counts were not significantly different for Bitcoin before and after the crypto rewards were available (t=-1.60). However, the donation counts were significantly higher for Ethereum when the crypto rewards became available (t=8.46). The average donation size dropped significantly for both Bitcoin and Ethereum (t=-2.60 for Bitcoin and t=-9.98 for Ethereum), and this drop is more substantial for Ethereum. Further, the past transaction number for the donated wallets with and without the airdrop did not drop significantly for BTC but significantly for ETH (t=-3.26), indicating selection.
The data trends are presented in Figures \ref{fig:donation_count} and \ref{fig:donation_avg}. These figures show parallel trends before the introduction of crypto rewards. From Figure \ref{fig:donation_count}, the hourly counts of donations followed similar trends prior to the announcement of a potential airdrop; the hourly donation counts on Ethereum increased after the announcement of a potential airdrop and peaked after the airdrop was confirmed. From Figure \ref{fig:donation_avg}, the average contribution size is higher for Ethereum before the introduction of the crypto rewards. After the crypto rewards were introduced, the average contribution size became similar between Bitcoin and Ethereum.

\subsection{Results}

We summarize the regression results for the aggregated analysis and transactional analysis in Table \ref{table:results}. Models 1 and 2 are the results of the aggregated analyses. Models 3 and 4 are the results of the transactional analysis without matching. Models 5 and 6 are the results of the transactional analysis with matched wallets. 

As can be seen from the coefficient of Airdrop × Ether of Models 1 and 2 in Table \ref{table:results}, the hourly donation count ($DonCount$) of the Ethereum blockchain (plus one) has increased 706.07\% more than the Bitcoin blockchain by the airdrop. We also learn from Model 2 of Table \ref{table:results} that the decrease in the average contribution sizes ($AvgDonSize$) is about 57\% more aggressive for Ethereum than for Bitcoin in response to the airdrop. This blockchain-level analysis does not account for the potential selection issue -- users who chose to donate via Ethereum may have a higher investment mindset and thus differ from those who donated via Bitcoin. As such, the blockchain-level analysis is appropriate for the outcome of $DonCount$ but not so much for the outcome of $AvgDonSize$.

We account for the selection issue and shed light on the outcome of donation sizes ($DonSize$) by performing matching at a wallet level and conducting transactional analyses. For the outcome of $DonSize$, we find consistent evidence from the negative and significant coefficient of Airdrop × Ether in Models 3, 4, 5, 6 that the crypto rewards likely led to a decrease in contribution sizes.

\begin{table} \centering 
  \caption{Results of Study I} 
    \renewcommand{\arraystretch}{1}
\begin{tabular}{c c c c c c c}
\hline 
\hline 
%\cline{2-7} 
 & \multicolumn{2}{c}{\textit{Aggregate (Hourly)}} &\multicolumn{2}{c}{\textit{Transactions}}& \multicolumn{2}{c}{\textit{Matched Transactions}} \\ 
 \hline
DV: &  $DonCount$  &  $AvgDonSize$  &  $DonSize$  &  $DonSize$ &  $DonSize$  &  $DonSize$  \\ 
& \multicolumn{1}{c}{(1)} & \multicolumn{1}{c}{(2)} & \multicolumn{1}{c}{(3)} & \multicolumn{1}{c}{(4)} & \multicolumn{1}{c}{(5)} & \multicolumn{1}{c}{(6)}\\ 
\hline 
  $Airdrop\times Ether$ &$2.087^{***}$  & $-0.451^{***}$& $-0.737^{***}$ & $-0.510^{***}$ & $-0.884^{***}$ & $-0.946^{***}$ \\ 
  &(0.164)&(0.119)& (0.038) & (0.038) & (0.093) & (0.091) \\ 
 $Airdrop$ &$3.736^{***}$ & $-0.138$& $0.249$ & $-0.118$ & $0.194$ & $-0.075$ \\ 
  &(0.625)&(0.557)& (1.082) & (1.058)  & (1.078) & (1.056) \\ 
  $Ether$ &$0.361$ & $0.073$& $-0.163^{***}$ & $-0.405^{***}$ & $-0.267^{**}$ & $-0.229^{**}$ \\ 
  &(0.236)&(0.172)& (0.036)& (0.036) & (0.113) &  (0.111)  \\ 
 $Fee Rate$ &$-0.007^{*}$ & $0.006^{**}$& $0.005^{***}$ & $0.006^{***}$ & $0.008^{***}$ & $0.007^{***}$ \\ 
  &(0.004)&(0.003)& (0.001) & (0.0005) & (0.002) & (0.002) \\ 
  $TimeSinceFirstSend$ &-- & --& -- & $-0.058^{***}$ & -- & $-0.181^{***}$ \\ 
  &--&--& --& $(0.015) $ & --&  (0.025) \\ 
    $TimeSinceFirstRec$ &-- & --& -- & $0.139^{***}$ & -- & $0.079^{***}$ \\ 
  &--&--& -- & $(0.014)$ & -- & $ (0.021)$ \\ 
    $Log(PastTrans+1)$ &-- & --& -- & $0.092^{***}$ & -- & $0.290^{***}$ \\ 
  &--&--&-- &$(0.002) $ & -- & (0.007) \\ 
  Intercept &$0.624$ & $5.082^{***}$& $3.676^{***}$ & $3.525^{***}$ & $3.736^{***}$ & $3.407^{***}$ \\ 
  &(0.443)&(0.458)& (1.080)  & (1.056) & (0.964) & (0.944) \\ 
 \hline 
   Time Dummy &Yes&Yes& \multicolumn{1}{c}{Yes} & \multicolumn{1}{c}{Yes} & \multicolumn{1}{c}{Yes} & \multicolumn{1}{c}{Yes} \\ \hline
Observations &268&267&  67,615  &  67,615  &  31,075  &  31,075  \\ 
R$^{2}$ &\multicolumn{1}{c}{0.887}&\multicolumn{1}{c}{0.686}& \multicolumn{1}{c}{0.158} & \multicolumn{1}{c}{0.195} & \multicolumn{1}{c}{0.128} & \multicolumn{1}{c}{0.163} \\ 
Adjusted R$^{2}$ &\multicolumn{1}{c}{0.769}&\multicolumn{1}{c}{0.357}& \multicolumn{1}{c}{0.143} & \multicolumn{1}{c}{0.181} & \multicolumn{1}{c}{0.124} & \multicolumn{1}{c}{0.159} \\ 
\hline 
\hline 
\textit{Note:}  & \multicolumn{6}{r}{$^{*}$p$<$0.1; $^{**}$p$<$0.05; $^{***}$p$<$0.01} \\ 
\end{tabular} 
\label{table:results}
\end{table}

\subsection{Donor Heterogeneity Analyses}
The negative link between crypto rewards and contribution size could vary by donor characteristics, and we perform moderation analyses to deepen our understanding of donor heterogeneity based on novel blockchain-based moderators on Ethereum.
\subsubsection{Ethereum Name Service.}
Cryptocurrency addresses represent long strings of numbers and letters, making it hard for one to send funds to another using Ethereum's networks. Ethereum Name Service is a distributed, open, and expandable naming system that maps human-readable Ethereum addresses to hexadecimal characters. For example, the machine-readable address of ``0x00d936ef12a4Fde33Ab0FcF08F18d6A9BAbB6b97'' would be translated into ``john.eth'' via ENS. ENS comes with an expiration date, and users need to pay for the continuous service at the annual rate of \$5 - \$30. ENS adopters use ENS IDs in various platforms as their identity - about 10\% of ENS adopters in our dataset use their ENS IDs as their Twitter handles. 

We examined the moderating role of ENS adoption on the relationship between crypto rewards and donation amounts by performing a transaction-level analysis within Ethereum. We identified a significantly positive moderating effect of ENS adoption. The details of the analysis are documented in Section 2 of the Appendix and discussed in a later section.
\subsubsection{Intermediary Platform Usage.}
People could donate to the Ukraine fundraising plea by directly sending funds to the Ethereum address of the Ukraine government (e.g., using MetaMask); they can also make donations using intermediary platforms such as Coinbase and Binance, which are both online platforms for buying, selling, transferring, and storing cryptocurrency. From the transactional data recorded in Etherscan, a direct donation's sender corresponds to an individual Ethereum address, while an indirect donation transaction's sender would be the intermediary platform of ``Coinbase'' or ``Binance.'' It is believed that direct transfers of funds would more likely make donors eligible for winning the crypto rewards, and the indirect transfers of funds via intermediary platforms would disqualify users from receiving the crypto rewards.\footnote{As anecdotal evidence, a \href{https://www.youtube.com/watch?v=dSJzSOOhEs4}{Youtuber} recommends not to use intermediary platforms for a higher chance of receiving the crypto rewards}. As reported in Section 2 of the Appendix, we identify a positive moderation effect of intermediary platform usage. 
\subsection{Robustness Checks}
To validate the results of our main analyses, we perform a battery of robustness checks. First, we removed extremely small-sized donations (i.e., donations less than \$5 or \$1) for both Bitcoin and Ethereum, and our findings remain unchanged (Appendix 3.1). Second, we altered the definition of treatment and allowed the treatment to end not on the scheduled snapshot but when it was cancelled. Our results remain unchanged (Appendix 3.2). Third, we change the observation window to include the time window of the potential airdrop, which we have removed in the main analysis, and our findings remain unchanged (Appendix 3.3). Last but not least, we keep only the first donation transaction of every wallet, and find our results unchanged (Appendix 3.4).

\section{Study II -- A Dictator Game in the Laboratory Setting}
Study I underscores the potential of crypto rewards to stimulate giving. However, it was based on an unprecedented fundraising event initiated by the Ukrainian government. The Ukrainian government also did not specify whether the crypto rewards would be fungible or non-fungible, while crypto rewards used in crypto donations are oftentimes NFTs \citep{liang2024market}. 

To assess the generalizability of the findings of Study I and to test our hypotheses H\ref{hyp:second} and H\ref{hyp:third}, we perform a laboratory experiment following a dictator game. Dictator games are considered a workhorse to understand charitable giving in both the economics and social psychology literature \citep{cartwright2023using,engel2011dictator,frey2001motivation}. In a dictator game, the dictator determines how to split an endowment between the self and the recipient. The dictator's action space is complete, ranging from giving nothing to giving everything, and the recipient has no influence on the endowment allocation. A dictator game has been used to examine various factors that influence giving behavior, such as rebate and donation matching, fairness considerations, social norms, and intrinsic motivations \citep{engel2011dictator, list2007interpretation}. \cite{list2007interpretation} underscored the importance of designing treatments to shed light on relevant field applications. We follow this direction to design a dictator game that invokes the most realistic responses for a set of highly relevant treatments concerning crypto rewards.

\subsection{Experimental Design}

\begin{figure}
    \centering
    \includegraphics[width=0.8\linewidth]{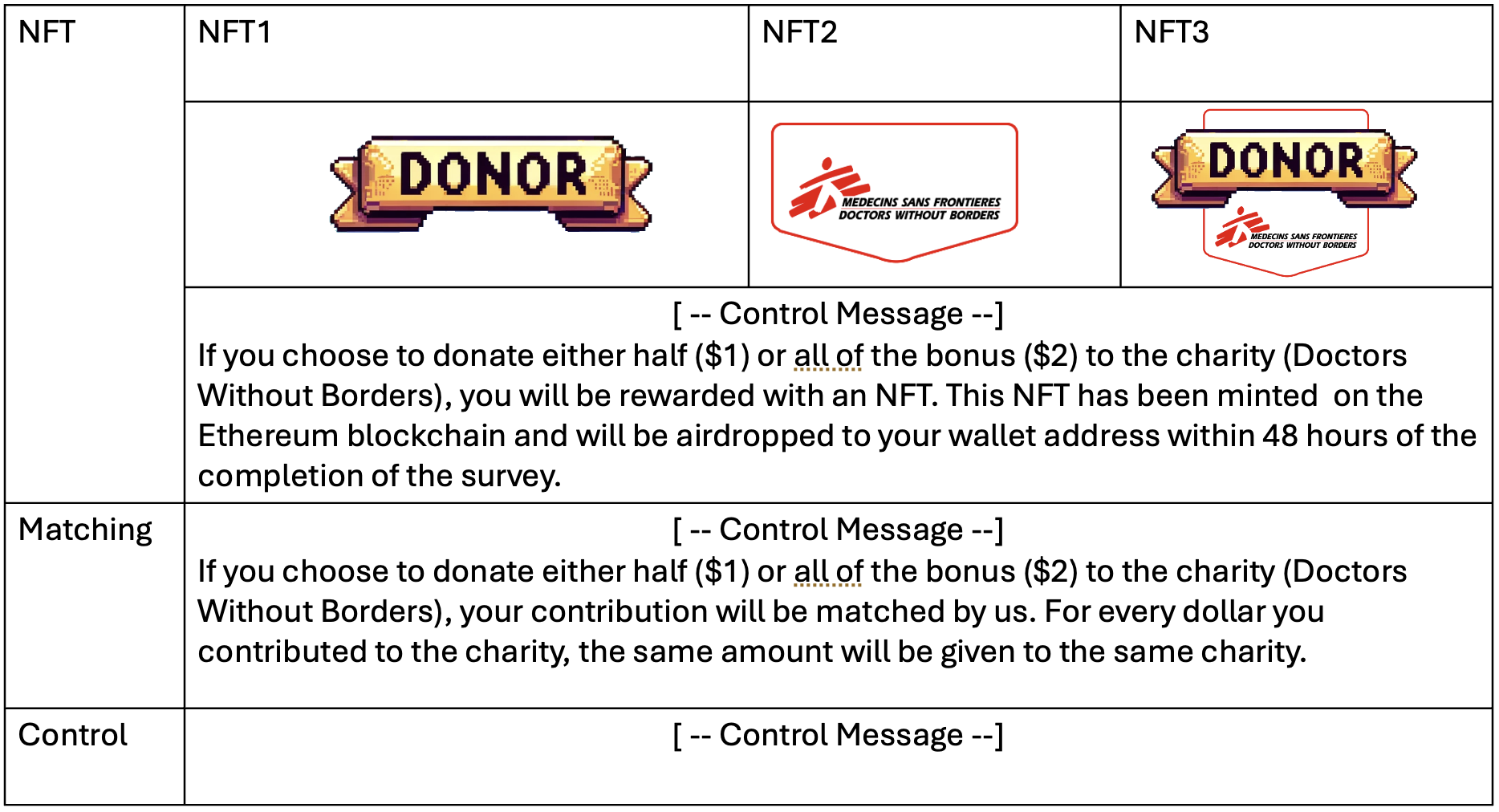}
    \caption{Experimental Design}
    \label{fig:experimental_groups}
\end{figure}

\begin{figure}
    \centering
    \includegraphics[width=0.8\linewidth]{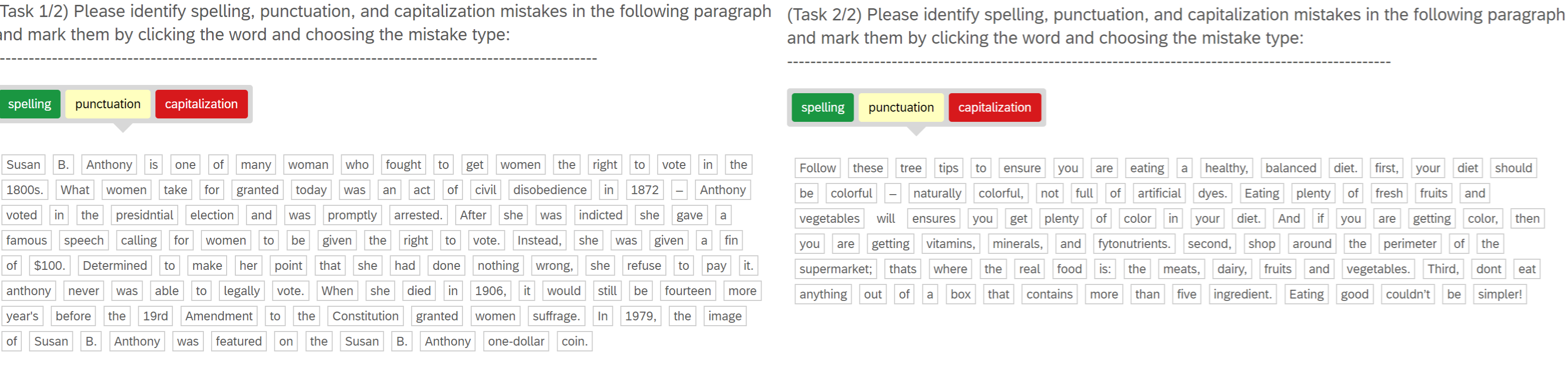}
    \caption{Study II - Tasks Before Income Allocation}
    \label{fig:tasks}
\end{figure}

We designed a between-subject experiment; it is preferred over a within-subject design, where subjects may be unable to distinguish different scenarios fully \citep{eckel2006subsidizing}. We recruited 286 subjects who have NFT experience
(self-disclosed as ``I own one or more NFTs, I have created one or more NFTs") from Prolific, a platform that allows researchers to recruit and manage participants for their online research. These subjects are randomly assigned to five groups: \textit{Control}, \textit{Matching}, \textit{NFT1}, \textit{NFT2}, and \textit{NFT3} (Figure \ref{fig:experimental_groups}). Past studies of dictator games reveal that dictators are more generous when the endowment is manna from heaven rather than earned by the dictators \citep{engel2011dictator}. To invoke realistic responses, subjects of our study are recruited to complete two copyediting tasks (Figure \ref{fig:tasks}). Specifically, subjects need to identify spelling, punctuation, and capitalization mistakes in two paragraphs of text to receive a flat-rate payment of \$2, on top of the payment of \$1 for completing the survey. The tasks are standard none-depletion tasks, and past literature has suggested a similar impact between a flat rate payment and a performance-based payment \citep{achtziger2015money}. After subjects are informed about their extra income for the copyediting tasks (\$2), the subjects are asked to choose among three options (the \textbf{control message}): (A) keep all the bonus (\$2) to themselves, (B) donate half of the bonus (\$1) to a charity (Doctors Without Borders) and keep the reminder, or (C) donate all the bonus (\$2) to the charity (Doctors Without Borders). This message is the control message available in all conditions. Doctors Without Borders works in over 70 countries to provide urgently needed humanitarian aid in moments of crisis. At the time of the experiment, multiple conflicts were taking place in the world, and Doctors Without Borders is a well-deserving charity. The fund allocation choices are semi-continuous, allowing us to assess whether to give (whether Option A is chosen) and how much to give (whether Option B or Option C is chosen) \citep{engel2011dictator}. If subjects are assigned to the Matching group, they receive an additional message that if they donate either \$1 or \$2 to the charity, their contribution will be matched. If they are assigned to any of the three NFT groups, they will receive an additional message with the visual design of the NFT reward that if they donate either \$1 or \$2 to the charity, they will receive the NFT reward. The NFT rewards have been minted on the Ethereum blockchain, and OpenSea links have been provided. However, these NFT rewards likely have no resell value as they were created by the experimenters. If subjects choose to donate, they also enter their wallet address so the NFT can be airdropped.

We included one attention check question -- participants were asked to indicate which charity was mentioned as a donation recipient; subjects were dropped if their answers were wrong. Out of the 286 subjects, 221 of them are male, 63 of them are female, one of them has a non-binary gender, and one prefers not to say. Regarding age, 39 of the subjects are between 18 and 24; 116 are between 25 and 34; 83 are between 35 and 44; 36 are between 45 and 54; Nine are between 55 and 64; Two of them are between 65 and 74, and one of them is above 74. In addition, we collected their birth country, enthusiasm about blockchain, optimism about NFT, and knowledge about cryptocurrency using a five-point Likert scale to control for their heterogeneity.

\subsection{Results}
Following the literature, we use a hurdle model to analyze subjects' responses \citep{engel2011dictator,breitmoser2013estimation}. A hurdle model assumes that the decision to make a positive contribution and the decision of how much to give, conditional on the decision to give, are two separate processes. It is composed of a binary decision (whether to give) using a logistic model and a continuous decision (how much to give) using OLS estimation for positive contributions, after adjusting for the truncation. It suits the dictator games well, where the majority of subjects choose to keep all the funds. We report the results in Table \ref{tab:studyII}. Model 1 did not include subject characteristics, and Model 2 included subjects' age, perceived optimism about NFT, blockchain literacy, gender, country, and enthusiasm about blockchain. From the first stage estimation (DV: $Donate$), we find from both Models 1 and 2 that only the matching treatment significantly improved the subjects' willingness to give. From the second stage estimation (DV: $Amount$), we find that conditional on the decision to give, NFT2 marginally improves the contribution size in Model 2, and NFT3 significantly improves the contribution size both in Models 1 and 2. In Model 2, the estimated $\beta$ for NFT2 is $\beta_{NFT2}=0.272$, and that for NFT3 is $\beta_{NFT3}=0.426$, indicating that NFT3 has a more salient impact on the decision of how much to give. Finally, we observe a significant estimation for $ln(\sigma)$, where $\sigma$ is the standard deviation parameter for the truncated normal distribution in the second stage (DV: $Amount$). 

\begin{table}
 \caption{Results of Study II}
 \renewcommand{\arraystretch}{1}
 \centering
\begin{tabular}{l*{2}{cc}}   
\hline\hline
            &\multicolumn{1}{c}{(1)}         &            &\multicolumn{1}{c}{(2)}         &            \\
\hline
DV: $Donate$&                     &            &                     &            \\
$Matching$     &       $0.615^{**}$ &     (0.240)&       $0.667^{***}$&     (0.253)\\
$NFT1$     &       0.157         &     (0.235)&       0.112         &     (0.245)\\
$NFT2$    &      -0.124         &     (0.257)&      -0.153         &     (0.276)\\
$NFT3$     &      -0.305         &     (0.262)&      -0.237         &     (0.269)\\ 
Constant      &      $-0.663^{***}$&     (0.166)&      -0.415         &     (1.011)\\ \hline
DV: $Amount$      &                     &            &                     &            \\
$Matching$     &       0.024         &     (0.135)&       0.026         &     (0.130)\\
$NFT1$     &       0.035         &     (0.144)&       0.133         &     (0.156)\\
$NFT2$    &       0.190         &     (0.166)&       $0.272^{*}$  &     (0.165)\\
$NFT3$     &       $0.383^{**}$ &     (0.176)&       $0.426^{**}$ &     (0.168)\\
Constant     &       $1.173^{***}$&     (0.105)&       $1.151^{**}$ &     (0.516)\\ \hline 
$ln(\sigma)$     &      $-0.855^{***}$&     (0.082)&      $-0.962^{***}$&     (0.081)\\  \hline

\multicolumn{1}{l}{Age/Optimism/Literacy} & No  & No  & Yes  & Yes \\
\multicolumn{1}{l}{Gender/Country/Enthusiasm} & No  & No  & Yes  & Yes \\
\hline
      Pseudo R2    &\multicolumn{1}{c}{0.048}    &   &\multicolumn{1}{c}{0.149}  &  \\ 

\hline
\(N\)       &         286         &            &         286         &            \\
\hline\hline
 \textit{Note:}  & \multicolumn{4}{r}{$^{*}$p$<$0.1; $^{**}$p$<$0.05; $^{***}$p$<$0.01} \\ 
\end{tabular}
 \label{tab:studyII}
\end{table}

From Study II, we find that the crypto rewards are no longer effective in stimulating the decision to give. This is in great contrast to Study I, and we discuss this difference in the next section. Further, matching treatment outperforms the NFT treatments only for the outcome of whether to donate but not for the outcome of how much to give. When subjects choose not to donate, donor identity is not self-relevant; only when they choose to donate does the donor's identity become relevant and have a positive impact on the decision of how much to give. Last but not least, Study II reveals that donor identity alone is not sufficient; the NFT treatment would only be significant when both donor identity and recipient identity are included in the graphic design.

\section{Conclusions}
\subsection{Summary of Findings}
We performed two studies to examine the impact of crypto rewards in fundraising campaigns. Study I leverages a quasi-experimental design to estimate the impact of crypto rewards issued by the Ukrainian government. Study II is a laboratory experiment following a dictator game, where the crypto rewards were NFTs issued by the experimenters. Combining the results from Studies I and II, we conclude that H\ref{hyp:first} is partially supported. As we present in Table \ref{tab:hypothesis_testing}, crypto rewards are effective in enticing giving when they offer sufficient monetary return or strongly present symbolic recognition, as in Study I but not in Study II. We went on to show that a traditional donation matching strategy strictly dominates the NFT crypto rewards created by the experimenters and could effectively stimulate the decision of whether to give, supporting H\ref{hyp:second}. Finally, leveraging the different graphic designs of the NFT thank-you gifts, we show that for subjects who decided to give and thus were relevant to donor identity, NFT rewards that highlight both donor identity and the recipient organization would effectively increase contribution sizes, conditional on giving. This result is in support of H\ref{hyp:third}. 

\begin{table}
\caption{Hypothesis Testing}
 \renewcommand{\arraystretch}{1}
    \centering
    \begin{tabular}{c|c|c}\hline \hline
        Hypotheses & Results  & Studies  \\  \hline
        H1 & Partially supported: effective in Study I but not in Study II & Studies I and II  \\ \hline
        H2 & Supported for the outcome of whether to donate & Study II  \\ \hline
        H3 & Supported   & Study II  \\ \hline \hline
         \end{tabular}
    \label{tab:hypothesis_testing}
\end{table}

\subsection{Reconciling the Results of Studies I and II}
Crypto rewards effectively motivate people to give in Study I but not in Study II. We provide two major reasons for these seemingly conflicting results. First, the crypto rewards in Study I likely offered higher returns. The blockchain community has been a strong supporter of Ukraine, with numerous crypto initiatives aimed at providing aid. For example, the Ukraine flag NFT associated with LOVE tokens from UkraineDAO was sold for 2173.6 ETH.\footnote{https://www.coindesk.com/tech/2022/03/02/ukrainian-flag-nft-raises-675m-for-countrys-war-efforts/} Similarly, the Avatar for Ukraine campaign, endorsed by the Minister of Digital Transformation of Ukraine, raised 12,656 ETH for medical aid for Ukrainian defenders.\footnote{https://www.avatarsforukraine.com/} Given these examples, it is reasonable to expect that the crypto rewards for the Ukrainian government's fundraising plea in Study I would have higher monetary value than the crypto rewards created by the experimenters in Study II. To support this mechanism, our moderation analysis in Study I shows that when donors contribute through intermediary platforms, where they are less influenced by direct monetary incentives, the reduction in contribution sizes was mitigated. 

Second, the crypto rewards in Study I offer stronger symbolic recognition than those in Study II. The fundraiser in Study I was the Ukrainian government, whereas in Study II, it was the experimenters. Crypto rewards granted by the Ukrainian government function as immutable badges, allowing donors to publicly display their generosity and social standing. These rewards are likely to invoke a heightened sense of doing good, enhanced reputation, and pride \citep{samek2017selective}. In contrast, the crypto rewards in Study II merely served as receipts for contributions of \$1 or \$2 to Doctors Without Borders, using bonuses earned from a copyediting task. Given the critical role of recognition in fundraising, the crypto rewards from Study I are more effective in encouraging giving than those from Study II \citep{recognitionartEstablishingBudgetfor}. Moreover, our moderation analysis showed that Ethereum users who adopted ENS were less likely to reduce their contribution sizes. This finding supports the symbolic recognition mechanism, as ENS adopters are likely more attuned to the value of recognition. Consequently, the more pronounced symbolic power of crypto rewards of Study I made it more effective than those in Study II. 
%There are other differences between Studies I and II. For example, the fundraising causes in Studies I and II are different in terms of urgency, overhead expenditure, and agency. While these differences could lead to the differential desirability between the fundraising causes, they do not offer explanations to why crypto rewards were effective for Study I but not Study II. 

Crypto rewards likely led to a contribution size reduction in Study I but could potentially increase the contribution size in Study II. Past studies show that people could reduce charitable contributions due to extrinsic incentives because they shift donors' attention away from their compassion to help and strengthen a cost-benefit mindset. \cite{chao2017demotivating} find that such extrinsic incentives need to be visually salient to take effect. In Study I, the crypto rewards received extensive media coverage and represented an unprecedented event for the crypto community. While they are not visually salient, they are socially salient and could occupy the prospective donors' attention. As such, this ``motivational crowding-out'' mechanism manifested in Study I but not in Study II. Further, the crypto rewards of Study I were not specified when the airdrop was announced. In contrast, the graphic design of NFT thank-you gifts was presented to the subjects in Study II. Donor identity prime is an effective device to stimulate giving \citep{kessler2018identity}, and this effect is more salient in Study II due to the presentation of the NFT designs.

\subsection{Theoretical and Practical Implications}

Our study began with a unique fundraising event and concluded with a general assessment of crypto rewards through a carefully designed dictator game, yielding several important implications. First, while the Ukrainian government's crypto fundraising plea was highly successful, it was largely unknown whether this success was reproducible and whether this novel thank-you gift could replace a donation matching strategy. Our study provides a nuanced understanding of the conditions under which crypto rewards can be a powerful tool in fundraising. The success of crypto rewards is contingent upon their ability to offer tangible value and symbolic significance. When these elements are present, crypto rewards could potentially surpass traditional donation-matching strategies, particularly in contexts where the fundraising cause resonates with the values of the blockchain community. However, in situations where the expected returns and symbolic power of crypto rewards are limited, donation matching remains a more reliable and effective strategy. Understanding these dynamics can help optimize the use of crypto rewards in future fundraising efforts, ensuring they are deployed in contexts where they can achieve the greatest impact.

Crypto rewards, as reflected in airdrops, have been widely used as a promotional strategy by blockchain-based projects \citep{li2021operation}; its effectiveness in supporting social causes that are not blockchain-based is unseen and barely understood. Our study suggests that ICO has a great potential to stimulate donations to support social causes that are not blockchain-based. To increase the societal impact of blockchain technology and accelerate the adoption of blockchain, the founders and designers of blockchain should consider applying crypto rewards to various social movements and activities. Our finding also indicates the necessity for blockchains to support airdrops to effectively improve fundraising performance when crypto rewards are present. Blockchain designers should also improve the design of blockchain-related platforms to better support airdrops. For example, currently, some airdrops can only be issued if a donor makes a direct transfer of funds to the recipient's wallet. Donors who use intermediary platforms (e.g., Coinbase) will not receive the airdrop due to technology limitations. Blockchain designers can work with intermediary platforms to better design and streamline the airdrop process.  

Last but not least, our study sheds light on the design of NFT thank-you gifts and offers two prescriptive suggestions. First, fundraisers should showcase the NFT thank-you gifts before the commencement of fundraising efforts. This is because the graphic design of the NFT thank-you gifts could play a role in the giving decision. Second, the thank-you gifts could be designed to focus on both the donor's identity and the prospective recipient's identity to reinforce the acts of giving.

% Appendix here
% Options are (1) APPENDIX (with or without general title) or
%             (2) APPENDICES (if it has more than one unrelated sections)
% Outcomment the appropriate case if necessary
%
% \begin{APPENDIX}{<Title of the Appendix>}
% \end{APPENDIX}
%
%   or
%
%\newpage
 %\begin{APPENDICES}

%All the potential in the world won't amount to much if research isn't cited correctly, though. Make sure you include complete citation information for your references, including publication or re-trieval dates for website citations, publication year and volume and issue numbers for journal articles, publisher names and locations for books, reports, and conference proceedings, and page numbers for eve-rything, but especially for direct quotes. For citations of unpublished work, you need to include the date of update, as well as the name and address of the organization that sponsored the work. Take a look at the reference section below to see how references should be formatted.

% Appendix here
% Options are (1) APPENDIX (with or without general title) or
%             (2) APPENDICES (if it has more than one unrelated sections)
% Outcomment the appropriate case if necessary
%
% \begin{APPENDIX}{<Title of the Appendix>}
% \end{APPENDIX}
%
%   or
%
\newpage
\setcounter{equation}{0}
\renewcommand{\theequation}{A\arabic{equation}}

\setcounter{table}{0}
\renewcommand{\thetable}{A\arabic{table}}

\section*{Appendix I -- Ordered Treatment Effect Identification}
In both the blockchain-level and the transaction-level analyses, we leverage a modified DiD analysis to identify the ordered treatment effect, or the average treatment effect on the treated (Ethereum donors), represented by  $ATET(EB|E)=E[Outcome^{E}_{1}-Outcome^{B}_{1}|E]$, where $E[Outcome^{d}_{t}]$ represents the expected outcome, with $d\in[B,E,0]$ and $t\in[0,1]$. We use $d=B$ to illustrate the treatment to get the crypto reward with a low probability, as in the Bitcoin blockchain; we use $d=E$ to illustrate the treatment to get crypto rewards with a high probability, as in the Ethereum blockchain; we use $d=0$ to illustrate the condition when the treatment of an airdrop has not been announced or has stopped. We use $t=1$ to denote the time when the airdrop was available and $t=0$ to denote the time when it is not available. This is equivalent to the local treatment effects discussed in \cite{angrist1995identification}. We can re-write this equation such that: 
\begin{equation}
\begin{split}
&ATET(EB|E)=E[Outcome^{E}_{1}-Outcome^{B}_{1}|E]
=E[Outcome^{E}_{1}|E]-E[Outcome^{B}_{1}|E].
\end{split}
\end{equation}

We observe $E[Outcome^{E}_{1}|E]$ but not $E[Outcome^{B}_{1}|E]$, and draw inferences from the Bitcoin blockchain by leveraging the strong parallel assumption that $E[Outcome^B_{1}|E]-E[Outcome^0_{0}|E]=E[Outcome^B_{1}|B]-E[Outcome^0_{0}|B]$. This assumption is an equal effect size assumption that likely holds in our context if ETH holders and BTC holders are equivalently sensitive to external rewards. Specifically, it is equivalent to saying that the spike we observe from ETH in response to the airdrop would also occur in BTC if the airdrop is more likely to be issued in BTC rather than ETH. We believe that this assumption holds because, from a blockchain standpoint, ETH and BTC are interchangeable. While BTC and ETH holders may hold different beliefs (e.g., about intervention and decentralization), the value of the airdrop to ETH holders and BTC holders should be similar both in terms of monetary incentive and symbolic recognition. Thus, we can re-write  $ATET(EB|E)$ such that:
\begin{equation}
\begin{split}
&ATET(EB|E)=E[Outcome^{E}_{1}-Outcome^{B}_{1}|E] \\
&=E[Outcome^{E}_{1}|E]-E[Outcome^0_{0}|E]-E[Outcome^B_{1}|B]+E[Outcome^0_{0}|B],
\end{split}
\end{equation}
where every component of the right side of Equation (2) is observed. Even if we believe that this assumption does not hold (e.g., ETH holders may react more aggressively to the airdrop), we can partially identify the ordered treatment effects as long as $E[Outcome^B_{0}|E]-E[Outcome^0_{0}|E]=E[Outcome^B_{0}|B]-E[Outcome^0_{0}|B]$. This common trend assumption is widely used in classic DiD designs and is highly likely to hold given the common pre-intervention trends we illustrated in the manuscript. \cite{fricke2017identification}
proves that with partial identification, we can interpret the estimates as the lower bound in the magnitude for the treatment effect. 

%We can validate this assumption by examining the pre-intervention trends in Figure 5 and Figure 6.\footnote{Other than the graphical evidence, we also validate the common trend assumption using an augmented DiD regression to allow group-specific linear time trends \citep{hansen2017have}. Specifically, we modify the model to the following:
%\begin{equation}
%\begin{split}
%&Outcome_{c,t}=\beta_{0}+\beta_{1} %Ether_{c} Airdrop_{t}
%+\beta_{2}Ether_{c}+\beta_{2}Airdrop_{t}+\%beta_{3}Ether_{c}
%+\beta_{4}(Ether_{c} \times t) \\
%&+\eta_{t}+\epsilon_{c,t}.
%\end{split}
%\end{equation}
%We run the model in Equation (4) and report the results in Appendix B. We show that our treatment effects are credible because they are not sensitive to the model specification.} 

The detailed econometric models for the blockchain-level and transaction-level analyses are presented in Equations (1) and (2) of the manuscript.

\section*{Appendix II -- Ethereum Moderation Analyses}
As reported in Section 3.5 of the manuscript, we explore the mechanism behind the reduction of contribution sizes following the announcement of crypto rewards in Study I. This is done by performing two moderation analyses using only donation transactions in Ethereum. The two moderators of our choice are Ethereum Name Service (ENS) adoption ($ENS$) and intermediary platform usage ($Intermediaries$). We only use Ethereum transactions for this analysis because these two moderators only exist on the Ethereum blockchain. Since the moderation analyses only used Ethereum transactions, we did not include hourly effects. It is possible that users who adopted ENS and used intermediary platforms have varying wealth levels. As such, we include a control variable, $NFTValue$, to account for the wealth effect. We report the results in Table A1.
\begin{table} [H] \centering 
  \caption{Moderation -ENS and Intermediaries} 
  \label{} 
    \renewcommand{\arraystretch}{1}
\begin{tabular} {c c c}
\hline 
\hline  
DV: & \multicolumn{1}{c}{Log($DonSize$)} & \multicolumn{1}{c}{Log($DonSize$)} \\ 

\hline  
  $Airdrop$ & $-0.530^{***}\,(0.016)$& $-0.493^{***}\,(0.015)$  \\ 
    $ENS$ & $0.644^{***}\,(0.025)$ & -- \\ 
     $Intermediaries$ & -- & $1.077^{***}\,(0.036)$ \\
      $Airdrop \times ENS$ & $0.088^{***}\,(0.032)$ & --  \\
  $Airdrop \times Intermediaries$ & -- & $0.330^{***}\,(0.057)$  \\
  $NFTValue$  & $0.024^{***} \,(0.0005)$& $0.033^{***}\,(0.0005)$  \\
  Intercept & $3.417^{***}\,(0.015)$ & $4.293^{***}\,(0.913)$ \\ 
  Hour FE & Yes & Yes \\ 
 \hline
Observations & \multicolumn{1}{c}{55,746} & \multicolumn{1}{c}{55,746}  \\ 
R$^{2}$ & \multicolumn{1}{c}{0.109} & \multicolumn{1}{c}{ 0.178 } \\ 
Adjusted R$^{2}$ & \multicolumn{1}{c}{0.109} & \multicolumn{1}{c}{ 0.176}  \\ 
\hline 
\hline  
\textit{Note:}  & \multicolumn{2}{r}{$^{*}$p$<$0.1; $^{**}$p$<$0.05; $^{***}$p$<$0.01} \\ 
\end{tabular} 
\label{table:moderation-ens}
\end{table}

\section*{Appendix III -- Robustness Checks for Study I}

\subsection*{Remove Minuscule Donations}
In this robustness check, we remove minuscule donations, which are usually small in size. The 5th percentile of contribution size is \$1.01 for ETH and \$4.94 for BTC. We perform two analyses, with one removing donations below \$5 and the other removing donations below \$1. The results are reported in Tables A2 and A3, where we find consistent findings.
\begin{table}[]
\centering 
 \caption{Robustness - Remove Minuscule Donations <= \$5} 
 \renewcommand{\arraystretch}{1}
\begin{tabular}{ccccc}
\hline
 & \multicolumn{2}{c}{\textit{Aggregate (Hourly)}} & \multicolumn{2}{c}{\textit{Matched Transactions}} \\ \hline
DV: & $DonCount$ & $AvgDonSize$ & $DonSize$ & $DonSize$ \\
$Airdrop\times Ether$ & 2.059*** & -0.398*** & -0.885*** & -0.941*** \\
 & (0.134) & (0.087) & (0.096) & (0.094) \\
$Airdrop$ & 1.933*** & 0.245 & 0.194 & 0.261 \\
 & (0.512) & (0.406) & (0.895) & (0.890) \\
$Ether$ & 0.225 & 0.305*** & 0.062 & 0.069 \\
 & (0.146) & (0.095) & (0.096) & (0.096) \\
$TimeSinceFirstSend$ & -- & -- &  -- & -0.237*** \\
 & -- & -- & --  & (0.039) \\
$TimeSinceFirstRec$ & -- & -- &  -- & 0.247*** \\
 & -- & -- &  -- & (0.037) \\
$Log(PastTrans+1)$ & -- & -- &  -- & 0.098*** \\
 & -- & -- &  -- & (0.008) \\
 $FeeRate$  &  -0.007***  &  0.002  &  0.004***  &  0.004**  \\
  &  (0.002)  &  (0.001)  &  (0.002)  &  (0.002)   \\
Intercept & 0.682* & 4.987*** & 3.526*** & 3.129*** \\
 & (0.362) & (0.333) & (0.790) & (0.787) \\ \hline
Time Dummy & Yes & Yes & Yes & Yes \\ \hline
Observations & 268 & 267 & 22,687 & 22,687 \\
R$^{2}$ & 0.912 & 0.692 & 0.079 & 0.089 \\
Adjusted R$^{2}$ & 0.820 & 0.370 & 0.073 & 0.083 \\ \hline
\textit{Note:} & \multicolumn{4}{r}{$^{*}$p$<$0.1; $^{**}$p$<$0.05; $^{***}$p$<$0.01}
\end{tabular}
\end{table}

\begin{table}[H]
 \caption{Robustness - Remove Minuscule Donations <= \$1} 
 \centering
 \renewcommand{\arraystretch}{1}
\begin{tabular}{ccccc}
\hline
 & \multicolumn{2}{c}{\textit{Aggregate (Hourly)}} & \multicolumn{2}{c}{\textit{Matched Transactions}} \\ \hline
DV: & $DonCount$ & $AvgDonSize$ & $DonSize$ & $DonSize$ \\
$Airdrop\times Ether$ & 2.162*** & -0.496*** & -0.773*** & -0.835*** \\
 & (0.144) & (0.093) & (0.088) & (0.086) \\
$Airdrop$ & 1.914*** & 0.240 & 0.117 & -0.124 \\
 & (0.546) & (0.434) & (1.009) & (0.992) \\
$Ether$ & 0.293* & 0.236** & -0.179* & -0.143 \\
 & (0.156) & (0.101) & (0.107) & (0.105) \\
$TimeSinceFirstSend$ & -- & -- & --  & -0.115*** \\
 & -- & -- &  -- & (0.023) \\
$TimeSinceFirstRec$ & -- & -- &  -- & 0.044** \\
 & -- & -- &  -- & (0.020) \\
$Log(PastTrans+1)$ & -- & -- &  -- & 0.241*** \\
 & -- & -- &  -- & (0.008) \\
\multicolumn{1}{l}{$FeeRate$} & \multicolumn{1}{l}{-0.007***} & \multicolumn{1}{l}{0.003*} & \multicolumn{1}{l}{0.007***} & \multicolumn{1}{l}{0.007***} \\
\multicolumn{1}{l}{} & \multicolumn{1}{l}{(0.002)} & \multicolumn{1}{l}{(0.002)} & \multicolumn{1}{l}{(0.002)} & \multicolumn{1}{l}{(0.002)} \\
Intercept & 0.661* & 5.030*** & 3.668*** & 3.368*** \\
 & (0.386) & (0.355) & (0.901) & (0.886) \\ \hline
Time Dummy & Yes & Yes & Yes & Yes \\ \hline
Observations & 268 & 267 & 29,423 & 29,423 \\
R$^{2}$ & 0.907 & 0.698 & 0.129 & 0.159 \\
Adjusted R$^{2}$ & 0.810 & 0.382 & 0.125 & 0.155 \\ \hline
\textit{Note:} & \multicolumn{4}{r}{$^{*}$p$<$0.1; $^{**}$p$<$0.05; $^{***}$p$<$0.01}
\end{tabular}
\end{table}
\newpage
\subsection*{Alternative Treatment Definition}
In our main analysis, we consider the observation window from Feb. 26, 2022 to March 4, 2022, with the treatment lasting from the official announcement of the airdrop till the planned snapshot. In this robustness check, we changed the ending point to the cancellation of the airdrop on March 3 at 6:00 AM, 2022. The exclusion of the post-cancellation period allows a conservative estimation of the treatment effect. As can be seen from the results in Table A4 of this appendix, our findings stay robust. 

\begin{table}[H]
 \caption{Robustness - Alternative Treatment Definition} 
  \centering
  \renewcommand{\arraystretch}{1}
\begin{tabular}{ccccc}
\hline
 & \multicolumn{2}{c}{\textit{Aggregate (Hourly)}} & \multicolumn{2}{c}{\textit{Matched Transactions}} \\ \hline
DV: & $DonCount$ & $AvgDonSize$ & $DonSize$ & $DonSize$ \\
$Airdrop\times Ether$ & 2.777*** & -0.576*** & -0.885*** & -0.941*** \\
 & (0.111) & (0.105) & (0.096) & (0.094) \\
$Airdrop$ & 3.533*** & -0.237 & 0.023 & -0.095 \\
 & (0.383) & (0.446) & (0.972) & (0.952) \\
$Ether$ & 0.320*** & 0.182* & -0.290** & -0.255** \\
 & (0.107) & (0.102) & (0.113) & (0.110) \\
$TimeSinceFirstSend$ & -- & -- &  -- & -0.181*** \\
 & -- & -- &   --& (0.025) \\
$TimeSinceFirstRec$ & -- & -- &   --& 0.079*** \\
 & -- & -- &   --& (0.021) \\
$Log(PastTrans+1)$ & -- & -- &  -- & 0.281*** \\
 & -- & -- &   --& (0.008) \\
\multicolumn{1}{l}{$FeeRate$} & \multicolumn{1}{l}{-0.007***} & \multicolumn{1}{l}{0.003*} & \multicolumn{1}{l}{0.008***} & \multicolumn{1}{l}{0.007***} \\
\multicolumn{1}{l}{} & \multicolumn{1}{l}{(0.002)} & \multicolumn{1}{l}{(0.002)} & \multicolumn{1}{l}{(0.002)} & \multicolumn{1}{l}{(0.002)} \\
Intercept & 0.638** & 5.081*** & 3.766*** & 3.441*** \\
 & (0.269) & (0.363) & (0.964) & (0.945) \\ \hline
Time Dummy & Yes & Yes & Yes & Yes \\ \hline
Observations & 268 & 267 & 31,075 & 31,075 \\
R$^{2}$ & 0.956 & 0.693 & 0.127 & 0.163 \\
Adjusted R$^{2}$ & 0.910 & 0.372 & 0.124 & 0.159 \\ \hline
\textit{Note:} & \multicolumn{4}{r}{$^{*}$p$<$0.1; $^{**}$p$<$0.05; $^{***}$p$<$0.01}
\end{tabular}
\end{table}
\newpage
\subsection*{Alternative Time Window}
In our main analysis, we removed transactions that happened after the initial announcement of ``no airdrop" and the subsequent announcement of an airdrop from the Ukrainian government because this period is associated with a possible airdrop. In this robustness check, we included transactions that occurred during this time window and considered this period as no airdrop. As we show in Table A5 of this appendix, our results remained unchanged.
\begin{table}[H]
 \caption{Robustness - Alternative Time Window}
\centering
  \renewcommand{\arraystretch}{1}
\begin{tabular}{ccccc}
\hline
 & \multicolumn{2}{c}{\textit{Aggregate (Hourly)}} & \multicolumn{2}{c}{\textit{Matched Transactions}} \\ \hline
DV: & $DonCount$ & $AvgDonSize$ & $DonSize$ & $DonSize$ \\
$Airdrop\times Ether$ & 2.029*** & -0.479*** & -0.751*** & -0.804*** \\
 & (0.160) & (0.096) & (0.090) & (0.089) \\
$Airdrop$ & 1.991*** & 0.194 & 0.069 & -0.178 \\
 & (0.629) & (0.463) & (1.082) & (1.061) \\
$Ether$ & 0.561*** & 0.156 & -0.371*** & -0.337*** \\
 & (0.172) & (0.104) & (0.108) & (0.106) \\
$TimeSinceFirstSend$ & -- & -- & -- & -0.200*** \\
 & -- & -- & -- & (0.023) \\
$TimeSinceFirstRec$ & -- & -- & -- & 0.113*** \\
 & -- & -- & -- & (0.020) \\
$Log(PastTrans+1)$ & -- & -- &--  & 0.281*** \\
 & -- & -- & -- & (0.008) \\
\multicolumn{1}{l}{$FeeRate$} & \multicolumn{1}{l}{-0.009***} & \multicolumn{1}{l}{0.003*} & \multicolumn{1}{l}{0.008***} & \multicolumn{1}{l}{0.007***} \\
\multicolumn{1}{l}{} & \multicolumn{1}{l}{(0.003)} & \multicolumn{1}{l}{(0.002)} & \multicolumn{1}{l}{(0.002)} & \multicolumn{1}{l}{(0.002)} \\
Intercept & 0.551 & 5.102*** & 3.856*** & 3.480*** \\
 & (0.444) & (0.379) & (0.967) & (0.948) \\ \hline
Time Dummy & Yes & Yes & Yes & Yes \\ \hline
Observations & 314 & 313 & 36,425 & 36,425 \\
R$^{2}$ & 0.876 & 0.657 & 0.118 & 0.153 \\
Adjusted R$^{2}$ & 0.748 & 0.301 & 0.114 & 0.149 \\ \hline
\textit{Note:} & \multicolumn{4}{r}{$^{*}$p$<$0.1; $^{**}$p$<$0.05; $^{***}$p$<$0.01}
\end{tabular}
\end{table}
\newpage
\subsection*{Multiple Donations of the Same Wallet}
It is possible for one wallet to be associated with multiple donation transactions. In our main analysis, we sum up donations from the same wallet if they were transacted in the same minute. In this robustness check, we only keep the first donation transaction for each wallet, and our results remain unchanged. It is notable that the observation for the transactional analysis has reduced due to the way we process the data. As we show in Table A6 of this appendix, our results stay unchanged.
\begin{table}[H]
 \caption{Robustness - Multiple Transactions from the Same Wallet}
 \centering
  \renewcommand{\arraystretch}{1}
\begin{tabular}{ccccc}
\hline
 & \multicolumn{2}{c}{\textit{Aggregate (Hourly)}} & \multicolumn{2}{c}{\textit{Matched Transactions}} \\ \hline
$DV:$ & $DonCount$ & $AvgDonSize$ & $DonSize$ & $DonSize$ \\ \hline
$Airdrop\times Ether$ & 2.231*** & -0.525*** & -0.843*** & -0.911*** \\
 & (0.151) & (0.099) & (0.094) & (0.092) \\
$Airdrop$ & 1.704*** & 0.304 & 0.347 & 0.114 \\
 & (0.577) & (0.465) & (1.086) & (1.064) \\
$Ether$ & 0.236 & 0.176 & -0.263** & -0.226** \\
 & (0.165) & (0.109) & (0.114) & (0.112) \\
$TimeSinceFirstSend$ & -- & -- & -- & -0.184*** \\
 & -- & -- & -- & (0.025) \\
$TimeSinceFirstRec$ & -- & -- & -- & 0.080*** \\
 & -- & -- & -- & (0.021) \\
$Log(PastTrans+1)$ & -- & -- & -- & 0.285*** \\
 & -- & -- & -- & (0.008) \\
\multicolumn{1}{l}{$FeeRate$} & \multicolumn{1}{l}{-0.008**} & \multicolumn{1}{l}{0.004*} & \multicolumn{1}{l}{0.008***} & \multicolumn{1}{l}{0.007***} \\
\multicolumn{1}{l}{} & \multicolumn{1}{l}{(0.002)} & \multicolumn{1}{l}{(0.001)} & \multicolumn{1}{l}{(0.002)} & \multicolumn{1}{l}{(0.002)} \\
Intercept & 0.693* & 5.068*** & 3.740*** & 3.415*** \\
 & (0.408) & (0.381) & (0.961) & (0.942) \\ \hline
Time Dummy & Yes & Yes & Yes & Yes \\ \hline
Observations & 268 & 267 & 30,802 & 30,802 \\
R$^{2}$ & 0.901 & 0.680 & 0.129 & 0.164 \\
Adjusted R$^{2}$ & 0.799 & 0.345 & 0.125 & 0.160 \\ \hline
\textit{Note:} & \multicolumn{4}{r}{$^{*}$p$<$0.1; $^{**}$p$<$0.05; $^{***}$p$<$0.01}
\end{tabular}
\end{table}

% etc

%\section*{Acknowledgments}
%This was was supported in part by......

%Bibliography
%\bibliographystyle{unsrt}  
\bibliographystyle{plainnat}
\bibliography{references}  

\begin{thebibliography}{43}
\providecommand{\natexlab}[1]{#1}
\providecommand{\url}[1]{\texttt{#1}}
\expandafter\ifx\csname urlstyle\endcsname\relax
  \providecommand{\doi}[1]{doi: #1}\else
  \providecommand{\doi}{doi: \begingroup \urlstyle{rm}\Url}\fi

\bibitem[Achtziger et~al.(2015)Achtziger, Al{\'o}s-Ferrer, and Wagner]{achtziger2015money}
Anja Achtziger, Carlos Al{\'o}s-Ferrer, and Alexander~K Wagner.
\newblock Money, depletion, and prosociality in the dictator game.
\newblock \emph{Journal of Neuroscience, Psychology, and Economics}, 8\penalty0 (1):\penalty0 1, 2015.

\bibitem[Akerlof and Kranton(2000)]{akerlof2000economics}
George~A Akerlof and Rachel~E Kranton.
\newblock Economics and identity.
\newblock \emph{The quarterly journal of economics}, 115\penalty0 (3):\penalty0 715--753, 2000.

\bibitem[Angrist and Imbens(1995)]{angrist1995identification}
Joshua Angrist and Guido Imbens.
\newblock Identification and estimation of local average treatment effects, 1995.

\bibitem[B{\'e}nabou and Tirole(2006)]{benabou2006incentives}
Roland B{\'e}nabou and Jean Tirole.
\newblock Incentives and prosocial behavior.
\newblock \emph{American economic review}, 96\penalty0 (5):\penalty0 1652--1678, 2006.

\bibitem[Breitmoser(2013)]{breitmoser2013estimation}
Yves Breitmoser.
\newblock Estimation of social preferences in generalized dictator games.
\newblock \emph{Economics Letters}, 121\penalty0 (2):\penalty0 192--197, 2013.

\bibitem[Briers et~al.(2007)Briers, Pandelaere, and Warlop]{briers2007adding}
Barbara Briers, Mario Pandelaere, and Luk Warlop.
\newblock Adding exchange to charity: a reference price explanation.
\newblock \emph{Journal of Economic Psychology}, 28\penalty0 (1):\penalty0 15--30, 2007.

\bibitem[Cartwright and Thompson(2023)]{cartwright2023using}
Edward Cartwright and Adam Thompson.
\newblock Using dictator game experiments to learn about charitable giving.
\newblock \emph{VOLUNTAS: International Journal of Voluntary and Nonprofit Organizations}, 34\penalty0 (1):\penalty0 185--191, 2023.

\bibitem[Chao(2017)]{chao2017demotivating}
Matthew Chao.
\newblock Demotivating incentives and motivation crowding out in charitable giving.
\newblock \emph{Proceedings of the National Academy of Sciences}, 114\penalty0 (28):\penalty0 7301--7306, 2017.

\bibitem[Chao and Fisher(2022)]{chao2022self}
Matthew Chao and Geoffrey Fisher.
\newblock Self-interested giving: The relationship between conditional gifts, charitable donations, and donor self-interestedness.
\newblock \emph{Management Science}, 68\penalty0 (6):\penalty0 4537--4567, 2022.

\bibitem[Duflo(2001)]{duflo2001schooling}
Esther Duflo.
\newblock Schooling and labor market consequences of school construction in indonesia: Evidence from an unusual policy experiment.
\newblock \emph{American economic review}, 91\penalty0 (4):\penalty0 795--813, 2001.

\bibitem[Eckel and Grossman(2003)]{eckel2003rebate}
Catherine~C Eckel and Philip~J Grossman.
\newblock Rebate versus matching: does how we subsidize charitable contributions matter?
\newblock \emph{Journal of Public Economics}, 87\penalty0 (3-4):\penalty0 681--701, 2003.

\bibitem[Eckel and Grossman(2006)]{eckel2006subsidizing}
Catherine~C Eckel and Philip~J Grossman.
\newblock Subsidizing charitable giving with rebates or matching: Further laboratory evidence.
\newblock \emph{Southern Economic Journal}, 72\penalty0 (4):\penalty0 794--807, 2006.

\bibitem[Engel(2011)]{engel2011dictator}
Christoph Engel.
\newblock Dictator games: A meta study.
\newblock \emph{Experimental economics}, 14:\penalty0 583--610, 2011.

\bibitem[Falk(2007)]{falk2007gift}
Armin Falk.
\newblock Gift exchange in the field.
\newblock \emph{Econometrica}, 75\penalty0 (5):\penalty0 1501--1511, 2007.

\bibitem[Felfe et~al.(2015)Felfe, Nollenberger, and Rodr{\'\i}guez-Planas]{felfe2015can}
Christina Felfe, Natalia Nollenberger, and N{\'u}ria Rodr{\'\i}guez-Planas.
\newblock Can’t buy mommy’s love? universal childcare and children’s long-term cognitive development.
\newblock \emph{Journal of population economics}, 28\penalty0 (2):\penalty0 393--422, 2015.

\bibitem[Frey and Jegen(2001)]{frey2001motivation}
Bruno~S Frey and Reto Jegen.
\newblock Motivation crowding theory.
\newblock \emph{Journal of economic surveys}, 15\penalty0 (5):\penalty0 589--611, 2001.

\bibitem[Frey and Oberholzer-Gee(1997)]{frey1997cost}
Bruno~S Frey and Felix Oberholzer-Gee.
\newblock The cost of price incentives: An empirical analysis of motivation crowding-out.
\newblock \emph{The American economic review}, 87\penalty0 (4):\penalty0 746--755, 1997.

\bibitem[Fricke(2017)]{fricke2017identification}
Hans Fricke.
\newblock Identification based on difference-in-differences approaches with multiple treatments.
\newblock \emph{Oxford Bulletin of Economics and Statistics}, 79\penalty0 (3):\penalty0 426--433, 2017.

\bibitem[Gandullia and Lezzi(2018)]{gandullia2018price}
Luca Gandullia and Emanuela Lezzi.
\newblock The price elasticity of charitable giving: New experimental evidence.
\newblock \emph{Economics Letters}, 173:\penalty0 88--91, 2018.

\bibitem[Gneezy and Rustichini(2000)]{gneezy2000fine}
Uri Gneezy and Aldo Rustichini.
\newblock A fine is a price.
\newblock \emph{The journal of legal studies}, 29\penalty0 (1):\penalty0 1--17, 2000.

\bibitem[Holmes et~al.(2002)Holmes, Miller, and Lerner]{holmes2002committing}
John~G Holmes, Dale~T Miller, and Melvin~J Lerner.
\newblock Committing altruism under the cloak of self-interest: The exchange fiction.
\newblock \emph{Journal of experimental social psychology}, 38\penalty0 (2):\penalty0 144--151, 2002.

\bibitem[Jung et~al.(2014)Jung, Nelson, Gneezy, and Gneezy]{jung2014paying}
Minah~H Jung, Leif~D Nelson, Ayelet Gneezy, and Uri Gneezy.
\newblock Paying more when paying for others.
\newblock \emph{Journal of personality and social psychology}, 107\penalty0 (3):\penalty0 414, 2014.

\bibitem[Kamas and Preston(2010)]{kamas2010can}
Linda Kamas and Anne Preston.
\newblock What can social preferences tell us about charitable giving? evidence on responses to price of giving, matching, and rebates.
\newblock In \emph{Charity with Choice}, pages 165--199. Emerald Group Publishing Limited, 2010.

\bibitem[Kessler and Milkman(2018)]{kessler2018identity}
Judd~B Kessler and Katherine~L Milkman.
\newblock Identity in charitable giving.
\newblock \emph{Management Science}, 64\penalty0 (2):\penalty0 845--859, 2018.

\bibitem[Kim et~al.(2020)Kim, Hong, Hwang, Kim, and Han]{kim2020comparison}
Hee~Jin Kim, Ji~Sun Hong, Hyun~Chan Hwang, Sun~Mi Kim, and Doug~Hyun Han.
\newblock Comparison of psychological status and investment style between bitcoin investors and share investors.
\newblock \emph{Frontiers in Psychology}, 11:\penalty0 502295, 2020.

\bibitem[Kohn(2008)]{kohn2008brighter}
Alfie Kohn.
\newblock \emph{The brighter side of human nature: Altruism and empathy in everyday life}.
\newblock Basic Books, 2008.

\bibitem[Li et~al.(2021)Li, Wan, Cheng, and Zhao]{li2021operation}
Jian Li, Xiang~Shawn Wan, Hsing~Kenneth Cheng, and Xi~Zhao.
\newblock Operation dumbo drop: To airdrop or not to airdrop for initial coin offering success?
\newblock \emph{Hsing Kenneth and Zhao, Xi, Operation Dumbo Drop: To Airdrop or Not to Airdrop for Initial Coin Offering Success}, 2021.

\bibitem[Liang et~al.(2024)Liang, Tunc, and Burtch]{liang2024market}
Chen Liang, Murat Tunc, and Gordon Burtch.
\newblock Market responses to genuine versus strategic generosity: An empirical examination of nft charity fundraisers.
\newblock \emph{arXiv preprint arXiv:2401.12064}, 2024.

\bibitem[List(2007)]{list2007interpretation}
John~A List.
\newblock On the interpretation of giving in dictator games.
\newblock \emph{Journal of Political economy}, 115\penalty0 (3):\penalty0 482--493, 2007.

\bibitem[List(2008)]{list2008introduction}
John~A List.
\newblock Introduction to field experiments in economics with applications to the economics of charity.
\newblock \emph{Experimental Economics}, 11:\penalty0 203--212, 2008.

\bibitem[List and Lucking-Reiley(2002)]{list2002effects}
John~A List and David Lucking-Reiley.
\newblock The effects of seed money and refunds on charitable giving: Experimental evidence from a university capital campaign.
\newblock \emph{Journal of political Economy}, 110\penalty0 (1):\penalty0 215--233, 2002.

\bibitem[Liu and Feng(2021)]{liu2021does}
Yuewen Liu and Juan Feng.
\newblock Does money talk? the impact of monetary incentives on user-generated content contributions.
\newblock \emph{Information Systems Research}, 32\penalty0 (2):\penalty0 394--409, 2021.

\bibitem[Miller(1999)]{miller1999norm}
Dale~T Miller.
\newblock The norm of self-interest.
\newblock \emph{American Psychologist}, 54\penalty0 (12):\penalty0 1053, 1999.

\bibitem[Miller and Prentice(1994)]{miller1994collective}
Dale~T Miller and Deborah~A Prentice.
\newblock Collective errors and errors about the collective.
\newblock \emph{Personality and Social Psychology Bulletin}, 20\penalty0 (5):\penalty0 541--550, 1994.

\bibitem[Newman and Shen(2012)]{newman2012counterintuitive}
George~E Newman and Y~Jeremy Shen.
\newblock The counterintuitive effects of thank-you gifts on charitable giving.
\newblock \emph{Journal of economic psychology}, 33\penalty0 (5):\penalty0 973--983, 2012.

\bibitem[Ramaswamy(2022)]{ramaswamy_2022}
Anita Ramaswamy.
\newblock Why web3's wealthy are donating crypto instead of cash, Mar 2022.
\newblock URL \url{https://techcrunch.com/2022/03/20/web3-charity-donate-crypto-cryptocurrency-nonprofit-cash/}.

\bibitem[{Recognition Art}(2023)]{recognitionartEstablishingBudgetfor}
{Recognition Art}.
\newblock {E}stablishing a {B}udget for {Y}our {D}onor {A}ppreciation {P}roject --- recognitionart.com.
\newblock \url{https://recognitionart.com/ideas/establishing-a-budget-for-your-donor-appreciation-project/}, 2023.
\newblock [Accessed 07-02-2024].

\bibitem[Samek and Sheremeta(2017)]{samek2017selective}
Anya Samek and Roman~M Sheremeta.
\newblock Selective recognition: How to recognize donors to increase charitable giving.
\newblock \emph{Economic Inquiry}, 55\penalty0 (3):\penalty0 1489--1496, 2017.

\bibitem[Steele(1997)]{steele1997threat}
Claude~M Steele.
\newblock A threat in the air: How stereotypes shape intellectual identity and performance.
\newblock \emph{American psychologist}, 52\penalty0 (6):\penalty0 613, 1997.

\bibitem[{The Giving Block}(2023)]{thegivingblock2023Annual}
{The Giving Block}.
\newblock {T}he 2023 {A}nnual {R}eport on {C}rypto {P}hilanthropy.
\newblock \url{https://thegivingblock.com/annual-report/}, 2023.
\newblock [Accessed 07-02-2024].

\bibitem[Velichety et~al.(2019)Velichety, Ram, and Bockstedt]{velichety2019quality}
Srikar Velichety, Sudha Ram, and Jesse Bockstedt.
\newblock Quality assessment of peer-produced content in knowledge repositories using development and coordination activities.
\newblock \emph{Journal of Management Information Systems}, 36\penalty0 (2):\penalty0 478--512, 2019.

\bibitem[Wing et~al.(2018)Wing, Simon, and Bello-Gomez]{wing2018designing}
Coady Wing, Kosali Simon, and Ricardo~A Bello-Gomez.
\newblock Designing difference in difference studies: best practices for public health policy research.
\newblock \emph{Annu Rev Public Health}, 39\penalty0 (1):\penalty0 453--469, 2018.

\bibitem[Zlatev and Miller(2016)]{zlatev2016selfishly}
Julian~J Zlatev and Dale~T Miller.
\newblock Selfishly benevolent or benevolently selfish: When self-interest undermines versus promotes prosocial behavior.
\newblock \emph{Organizational Behavior and Human Decision Processes}, 137:\penalty0 112--122, 2016.

\end{thebibliography}

\end{document}